\documentclass[letterpaper,12pt]{article}
\usepackage{graphicx}
\usepackage{color,amssymb,enumerate,float,amsmath}
\usepackage{tikz}
\usetikzlibrary{matrix}
\usepackage{caption}
\usepackage{subcaption}
\usepackage{ifpdf}
\ifpdf
	\usepackage[pdftex,unicode,implicit]{hyperref}

	\hypersetup{
  	pdftitle     = {}, 
  	pdfkeywords  = {},
  	pdfauthor    = {},
  	pdfcreator   = {pdf\LaTeXe\ with package \flqq hyperref\frqq},
  	pdfproducer  = {pdf\LaTeXe\ with package \flqq hyperref\frqq},
  	pdfpagemode  = UseNone,  
  	pdffitwindow = true,  
  	unicode      = true,
  	plainpages   = true,
  	colorlinks   = true,  
  	citecolor    = black,  
  	urlcolor     = blue, 
  	linkcolor    = black
	}

\else

  \usepackage[dvips]{graphicx}

\fi 

\makeatletter
\@addtoreset{equation}{section}
\makeatother

\begin{document}
	
\thispagestyle{empty}

\begin{center}
{\bf \LARGE Casimir effect of a rough membrane in 2+1 Ho\v rava-Lifshitz theory}
\vspace*{15mm}

{\large Claudio B\'orquez}$^{1,a}$
{\large and Byron Droguett}$^{2,b}$
\vspace{3ex}

$^1${\it Department of Physics, Universidad de Antofagasta, 1240000 Antofagasta, Chile.
}
$^2${\it Departamento de Ciencias B\'asicas, Facultad de Ciencias, Universidad Santo Tom\'as, Sede Arica, Chile.}

\vspace{3ex}

$^a${\tt cl.borquezg@gmail.com,}\hspace{.5em}
$^b${\tt byrondroguett@santotomas.cl}

\vspace*{15mm}
{\bf Abstract}
\begin{quotation}{\small\noindent
}
We investigate the Casimir effect of a rough membrane within the framework of the Ho\v rava-Lifshitz theory in $2+1$ dimensions. Quantum fluctuations are induced by an anisotropic scalar field subject to Dirichlet boundary conditions. We implement a coordinate transformation to render the membrane completely flat, treating the remaining terms associated with roughness as a potential. The spectrum is obtained through perturbation theory and regularized using the $\zeta$--function method. We present an explicit example of a membrane with periodic border. Additionally, we consider the effect of temperature. Our findings reveal that the Casimir energy and force depend on roughness, the anisotropic scaling factor and temperature.
\end{quotation}
\end{center}

\thispagestyle{empty}

\newpage

\section{Introduction}

The Casimir effect is a physical phenomenon predicted by quantum field theory that manifests itself through quantum fluctuations in the vacuum. The effect was first predicted theoretically by Casimir, who demonstrated that two parallel conducting and uncharged plates, separated by a distance significantly smaller than their dimensions, generate an attractive force between them \cite{Casimir:1948dh}. Subsequent studies have explored additional geometries, showing that the Casimir effect is sensitive to specific geometric configurations \cite{Boyer:1968uf}. This effect has been experimentally confirmed with a high degree of accuracy \cite{Lamoreaux:1996wh, Bressi:2002fr}.
The nature of the Casimir force has been discussed in various contexts because different factors interfere in this effect, including boundary conditions, topologically non-trivial backgrounds and temperature \cite{Bordag:2001qi, Teo:2011kt, Zhao:2006rr}. Certainly, all these characteristics modify the Casimir effect and cannot be neglected. The dimensionality of spacetime plays a fundamental role in the Casimir effect. In particular, research in low dimensions is essential for understanding the behavior of quantum fluctuations generated by various classes of fields on types of two-dimensional materials as boundaries, such as graphene family (for a comprehensive introduction, see \cite{Bellucci:2019ybj}). Knowledge of their properties will contribute significantly to advances in materials science.

Several different methods, such as the $\zeta$--function regularization method, have been developed to calculate the Casimir effect \cite{Kirsten:2010zp}. This method allows us to regulate the spectral sum over all eigenvalues associated with a particular operator. When considering the finite temperature case, it is often helpful to approach the problem within the framework of quantum field theory through an effective action, which can be expressed in terms of a $\zeta$--function.
In this research, our focus is on the study of the low dimensional Casimir effect of a rough membrane embedded in a flat $2+1$--dimensional manifold that satisfies Dirichlet boundary conditions, where the effects of temperature are considered.

Our objective is to investigate the Casimir effect in theories characterized by the breaking of Lorentz symmetry, particularly in Ho\v rava-Lifshitz-like theories. Different studies have addressed the Lorentz symmetry breaking in this context \cite{Anselmi:2008bq, Anselmi:2008bs}. Several cases have been investigated, which exhibit anisotropic behavior, including extensions involving Klein-Gordon and fermionic fields \cite{Ferrari:2010dj, MoralesUlion:2015tve, Muniz:2014dga, daSilva:2019iwn, Erdas:2023wzy}. Other research on Lorentz violation has incorporated terms into the Lagrangian with a preferred direction \cite{Cruz:2017kfo, Erdas:2020ilo}. Furthermore, finite temperature effects in these theories have been explored \cite{Cruz:2018bqt, Erdas:2021xvv, Cheng:2022mwd}. In a previous study, the Casimir effect has also been studied in the framework of the Ho\v rava gravity theory \cite{Borquez:2023cuf}. The Ho\v rava gravity emerges as a viable candidate to complement the high energy regime of General Relativity \cite{Horava:2009uw, Horava:2008ih, Blas:2009qj}. This theory is grounded in an anisotropic scaling of spacetime, leading to a reduction in general diffeomorphisms and a breaking of Lorentz symmetry in the ultraviolet. Reduction in symmetry gives rise to the appearance of an instantaneous scalar mode that persists even in 2+1 dimensions, in contrast to General Relativity. Therefore, the Ho\v rava theory presents a truly localized gravitational framework in $2+1$ dimensions. This physical mode is the response to carry gravitational interaction in the quantum formulation, even when dealing with a vacuum and the absence of a cosmological constant \cite{Horava:2009uw,Bellorin:2019zho}. The foliation of the theory is defined as a set of spatial hypersurfaces accompanied by a preferred temporal direction, which must be preserved by reduced diffeomorphisms. The most appropriate variables to describe this preferred foliation are the ADM variables \cite{Arnowitt:1962hi}. In Ref. \cite{Borquez:2023cuf}, we computed the Casimir effect resulting from the quantum fluctuations of an anisotropic scalar field on a membrane embedded in a conical manifold in 2 + 1 dimensions, induced by the presence of a massive point particle at rest located at the origin of the coordinates system. Consequently, the Casimir energy and force depend on the presence of the massive particle that generates the conical manifold, the anisotropic scaling factor and the temperature. All these considerations make it interesting to continue investigating the Casimir effect in low dimensions in the context of Ho\v rava-Lifshitz theories.

In this research, our primary focus is to investigate the Casimir effect resulting from the quantum fluctuations of an anisotropic scalar field acting on a rough membrane embedded in a flat $2+1$-dimensional manifold, which satisfy  Dirichlet boundary conditions. For a realistic application, we consider roughness as a perturbation from the flat case. Corrugated surfaces with diverse geometries have also been studied \cite{Kiani:2012gh, Setare:2014eva, Setare:2013vba}. To address the non-flat border, we initially implement a change of coordinates, ensuring that, in the new variables, the membrane assumes a completely flat border, and where the remaining terms associated with the roughness are incorporated into the potential \cite{Droguett:2014gqa}. To find the spectrum of eigenvalues, we employ perturbation theory up to first order and utilize the $\zeta$--function in the regularization process. Furthermore, we present a specific example of a  membrane with periodic border. Additionally, we take into consideration the impact of temperature through the effective action.

This paper is organized as follows. In section 2, we present the problem of the rough membrane and the solution to the eigenvalue problem through the perturbation theory. In section 3, we apply the regularization method using the $\zeta$--function and determine the energy and force density in the limit of infinite length. We present an explicit example of a membrane with periodic border. In section 4, we consider the effects of temperature on the membrane. Finally, in section 5, we present our conclusions.



\section{Rough membrane in Ho\v rava-Lifshitz theory}

The Ho\v rava gravity theory emerges as a candidate to complete the ultraviolet regime of General Relativity. This theory is grounded in an anisotropy between space and time
\begin{equation}
	[t]=-z\,,
 \qquad
 [x^i]=-1
 \,,
\end{equation}
where $z$ represents the anisotropy scaling factor. This leads to a reduction in the symmetry of general diffeomorphisms and a breaking of Lorentz symmetry in the ultraviolet sector. Consequently, the symmetry that preserves the foliation  takes the form
\begin{equation}
\delta t=f(t)\,,\qquad x^i=\zeta^i(t,x^k)\,,
\end{equation}
where the time is reparameterized on itself. Then the foliation has absolute physical meaning. The most suitable variables to describe this foliation are the ADM variables. This gives rise to the appearance of an additional scalar mode, instantaneous and even present in $2+1$ dimensions, making the theory in this dimension non-topological. The anisotropic scaling permits the inclusion of new terms in the Lagrangian density, which remain invariant under this scaling. Following this concept, it is possible to generalize field theories such as the Klein-Gordon, electrodynamic, and Yang-Mills theories. In this present study, quantum fluctuations are generated by the anisotropic scalar field, whose Lagrangian density is a generalization of the Klein-Gordon Lagrangian
\begin{eqnarray}
    S
    =
    \frac{1}{2}\int\,dt\,d^{2}xN\sqrt{g}
    \left(
    \partial_t\phi\partial_t\phi
    - l^{2(z-1)}\nabla_{i1}\cdots\nabla_{iz}\phi\nabla^{i1}\cdots\nabla^{iz}\phi
    \right)\,,
\end{eqnarray}
and the equation of motion for the scalar field is
\begin{equation}\label{KGeq}
(\partial_t^2+(-1)^{z}l^{2(z-1)}\nabla_{i1}\cdots\nabla_{iz}\nabla^{i1}\cdots\nabla^{iz})\phi=0\,,
\end{equation}
where the parameter $l$ has dimension of the inverse of mass. In general, the covariant derivative takes the form
\begin{eqnarray}\label{OLB}
    \nabla_{i1}\cdots\nabla_{iz}\nabla^{i1}\cdots\nabla^{iz}=\Delta^{z}\,.
\end{eqnarray}

In this research, we will compute the Casimir effect produced by a membrane with non-flat border embedded in a $2+1$-dimensional anisotropic spacetime. The membrane is modeled by the following coordinates
\begin{equation}
0\leq x\leq L\, \qquad 0\leq y\leq a+h(x)\,,
\end{equation}
where $L$ denotes the length of the membrane, and $a$ represents its width, with the assumption that $a\ll L$. Additionally, $h(x)$ characterizes the roughness of the membrane, satisfying the condition $h(x)\ll a$. To address the roughness, we implement a change of variables in such a way that, in the new variables, the membrane exhibits a flat border. The residual terms associated with the roughness are treated as a potential in the eigenvalue problem. Then, we consider the coordinates $x$ and $y=\rho\left(1 + h(x)/a\right)$, where
\begin{equation}
      0\leq x\leq L\,, \qquad 0\leq \rho\leq a\,,
\end{equation}
hence we can formulate the spatial metric in the new variables
\begin{equation}
g_{ij}=
\begin{pmatrix}
    1+
    \left(
    \frac{h^{'}\rho}{a}
    \right)^2 
    &
    \left(
    1+\frac{h}{a}
    \right) \frac{h^{'}\rho}{a} 
    \\
   \left(
   1+\frac{h}{a}
   \right) \frac{h^{'}\rho}{a} 
   & 
   \left(
   1+\frac{h}{a}
   \right)^2
   \\
\end{pmatrix}
\,.
\end{equation}
Since our aim is to solve an eigenvalue problem, we compute the Laplace-Beltrami operator
\begin{eqnarray}
    \Delta
    &=&
    \partial_x^{2}
    +\left[
    \frac{a^2+\left(\rho h^{'}
    \right)^2}{(a+h)^2}
    \right]\partial_\rho^2
    -\frac{2\rho h^{'}}{a+h}\partial_{x\rho}^2
    +\left[
    -\frac{\rho h^{''}}{a+h}
    +\frac{2\rho
    \left(
    h^{'}
    \right)^2}{(a+h)^2}
    \right]\partial_\rho
    \,.
\label{Beltrami}
\end{eqnarray}
We expand the operator (\ref{Beltrami}) in terms of $h(x)/a$ up to the second order in perturbations with the aim of treating the roughness as a potential.
Moreover, in order to operate with dimensionless coordinates we implement the following parameterization:
\begin{eqnarray}
\begin{split}
    x & = uL\,, \qquad 0\leq u\leq 1\,,
    \\
    \rho & = va\,,\qquad 0\leq v\leq 1\,.
\end{split}
\end{eqnarray}
Since we consider $L\gg h(x)$, we can discard several terms from the operator in the new coordinates, leading to a helpful simplification, then the Laplace-Beltrami operator has the new form
\begin{eqnarray}\label{BelSim}
    \Delta=
    \frac{1}{L^2}\partial_u^{2}
    +\frac{1}{a^2}\left[
    1
    -\frac{2\tilde{h}}{a}
    +3\left(
    \frac{\tilde{h}}{a}
    \right)^2
    \right]\partial_v^2
    \,,
    \end{eqnarray} 
with $\tilde{h}=h(uL)$.
Once the Laplace-Beltrami operator is obtained, using Eqs. (\ref{KGeq}) and (\ref{OLB}), we can define a spatial anisotropic operator $\mathcal{P}$ associated with the eigenvalue $\omega$, which depends on the scaling factor $z$
    \begin{equation}\label{POperator}
       \mathcal{P}\phi
       =
       (-1)^zl^{2(z-1)}\Delta^{z}\phi=\omega\phi\,.
    \end{equation}
The anisotropic scalar field satisfies  the  following Dirichlet boundary conditions:
\begin{eqnarray}
\begin{split}
        \phi(u,0)& =\phi(u,1) = 0\,,
        \\
        \phi(0,v) & =\phi(1,v) = 0\,.
\end{split}
\end{eqnarray}
In accordance with the structure of the anisotropic operator $\mathcal{P}$ in Eq. (\ref{POperator}), it is sufficient to calculate the eigenvalues of the Laplace-Beltrami operator defined in (\ref{BelSim}). To achieve this, we denote $\lambda$ as the eigenvalues associated with this operator. Consequently, the eigenvalue $\omega$ can be expressed as follows
\begin{eqnarray}\label{AutovalorOmega}
    \omega_{n,m}
    =
    l^{2(z-1)}
    \lambda_{n,m}^z\,.
\end{eqnarray}
To solve this eigenvalue problem in the presence of a rough membrane, we choose to use the perturbation theory.
For the zeroth order in perturbations, the partial differential equation is
    \begin{eqnarray}\label{EcDIffP}
        -\Delta\phi^{(0)} 
        = 
        -\left(\frac{1}{L^2}\partial_u^{2}
    +\frac{1}{a^2}\partial_v^2\right)\phi^{(0)}=\lambda^{(0)}\phi^{(0)}
    \,.
    \end{eqnarray}
The normalized eigenfunction associated with the Eq. (\ref{EcDIffP}) corresponds to
\begin{eqnarray}
       \phi^{(0)}_{n,m}(u,v)
       =
       2\sin(n\pi v)\sin(m\pi u)
       \,,
\end{eqnarray}
and hence its eigenvalues are
\begin{eqnarray}
        \lambda_{n,m}^{(0)}
        =
        \left(\frac{n\pi}{a}\right)^2+\left(\frac{m\pi}{L}\right)^2
        \,.
\end{eqnarray}    
To find the first order eigenvalues in perturbation theory, we must calculate the following integral
    \begin{eqnarray}
        \lambda_{n,m}^{(1)}
        =
        \int_0^1\int_0^1\,du\,dv\phi^{(0)*}_{n,m}(u,v)\mathcal{M}(u)\partial^2_{v}\phi^{(0)}_{n,m}(u,v)
        \,,
        \end{eqnarray}
where 
    \begin{eqnarray}
       \mathcal{M}(u)
       =
       \frac{1}{a^2}\left( \frac{2\tilde{h}}{a}
    -3\left(
    \frac{\tilde{h}}{a}
    \right)^2\right) 
    \,.
    \end{eqnarray}
Therefore, the total eigenvalues up to the first order in perturbation theory, are given by      
    \begin{eqnarray}\label{TotalEV}
        \lambda_{n,m}=
        \left(
        \frac{n\pi}{a}
        \right)^2
        +\left(
        \frac{m\pi}{L}
        \right)^2
        -(n\pi)^2\int_0^1\,du\mathcal{M}(u)
        +(n\pi)^2\int_0^1\,du\cos(2m\pi u)\mathcal{M}(u)
        \,.
        \nonumber
        \\
    \end{eqnarray}
Despite the fact that our method is applicable to any roughness $h(x)$, from this point forward, we will consider only periodic functions. Thus, the integration of the last term in the Eq. (\ref{TotalEV}) is zero.


\section{Casimir effect: $\zeta$--function regularization}

The vacuum expectation value of the energy can be regularized using the $\zeta$--function method associated with the eigenvalues (\ref{AutovalorOmega}) of the anisotropic spatial operator $\mathcal{P}$ \cite{Bordag:2001qi, Kirsten:2010zp}
\begin{eqnarray}\label{ZetaP}
    \zeta_{\mathcal{P}}(s)
    =
    l^{-2(z-1)s}\sum_{n,m\in\mathbb{N}}\left[
    \left(
    \frac{n\pi}{a}\right)^2
    +\left(
    \frac{m\pi}{L}
    \right)^2
        -(n\pi)^2\int_0^1\,du\mathcal{M}(u)
        \right]^{-sz}\,.
\end{eqnarray}
This function has the structure of the Epstein $\zeta$--function, then the integral form of the Eq. (\ref{ZetaP}) is
\begin{eqnarray}
    \zeta_{\mathcal{P}}(s)
    =
    \frac{l^{-2(z-1)s}}{\Gamma(sz)}
    \int_0^\infty\,dt\, t^{sz-1}
    \sum_{n,m\in\mathbb{N}}\exp\left[-t\left(r_1n^2+r_2m^2\right)\right],
\end{eqnarray}
where 
\begin{eqnarray}\label{R12}
    r_1
    =
    \pi^2\left(
    \frac{1}{a^2}-\int_0^1\,du\mathcal{M}(u)
    \right)
    \,,
    \qquad
    r_2 =
    \left(
    \frac{\pi}{L}
    \right)^2\,.
\end{eqnarray}
A suitable representation of the $\zeta$--function is obtained by using the Poisson resummation \cite{Kirsten:2010zp}
\begin{eqnarray}
    \zeta_{\mathcal{P}}(s)
    &=&
    \frac{l^{-2(z-1)s}}{4\Gamma(sz)}\left\{\int_0^\infty\,dt\,t^{sz-1} 
    + \frac{2\pi}{\sqrt{r_{1}r_{2}}}\int_0^\infty\,dt\,t^{sz-2}
    \left[\frac{1}{2} + \sum_{n=1}^{\infty}e^{\frac{-\pi^{2}n^{2}}{tr_{1}}} + \sum_{m=1}^{\infty}e^{\frac{-\pi^{2}m^{2}}{tr_{2}}}\right.\right.
    \nonumber
    \\
    &&
    \left.\left.
    + 2\sum_{n,m=1}^{\infty}e^{-\frac{\pi^{2}}{t}\left(\frac{n^{2}}{r_{1}}+\frac{m^{2}}{r_{2}}\right)}\right]
    - 2\sqrt{\pi}\int_0^\infty\,dt\,t^{sz-3/2}\left[\frac{1}{2}\left(\frac{\sqrt{r_1}+\sqrt{r_2}}{\sqrt{r_1r_2}}\right)\right.\right.
    \nonumber
    \\
    &&
    \left.\left.
    + \frac{1}{\sqrt{r_{1}}}\sum_{n=1}^{\infty}e^{\frac{-\pi^{2}n^{2}}{tr_{1}}}
    + \frac{1}{\sqrt{r_{2}}}\sum_{m=1}^{\infty}e^{\frac{-\pi^{2}m^{2}}{tr_{2}}}
    \right]\right\}
    \,.
    \label{zetaPs}
\end{eqnarray}
Our purpose is to determine the Casimir energy and force density, taking into account a rough membrane. To accomplish this, we must evaluate the case where $s=-1/2$ and concentrate on the finite component of the integral representation of the $\zeta$--function. Consequently, the Casimir energy is expressed as $E_{C}=\frac{1}{2}\zeta_{\mathcal{P}}(-1/2)$, hence we have
\begin{eqnarray}
    E_{C}
    &=&
    \frac{l^{z-1}}{4\Gamma(-z/2)}\left\{\pi^{-z-1}\Gamma(1+z/2)\zeta_{R}(2+z)\left(\frac{r_{1}^{\frac{1}{2}(1+z)}}{\sqrt{r_{2}}}+\frac{r_{2}^{\frac{1}{2}(1+z)}}{\sqrt{r_{1}}}\right)\right.
    \nonumber
    \\
    &&
    + 2\pi^{-z-1}(r_{1}r_{2})^{\frac{1}{2}(1+z)}\Gamma(1+z/2)\sum_{m,n=1}^{\infty}\left(r_{2}n^{2} + r_{1}m^{2}\right)^{-1-z/2}
    \nonumber
    \\
    &&
    \left.- \pi^{-z-1/2}\Gamma\left(\frac{1+z}{2}\right)\zeta_{R}(1+z)\left(r_{1}^{z/2} + r_{2}^{z/2}\right)\right\}
    \,.
\end{eqnarray}
We can notice that when the value of the anisotropic scaling factor $z$ takes on an even number, it causes the $\zeta$--function to tend to zero. Therefore, the Casimir energy and force vanish for these values.

To find the energy and force density we must divide by the length $L$. When $L$ tends to infinity it reduces significantly, hence the energy and force density are given, respectively, by
\begin{eqnarray}
    \mathcal{E}_{C}
    &=&
    l^{z-1}\frac{\Gamma(1+z/2)\zeta_{R}(2+z)}{4\pi^{z+2}\Gamma(-z/2)}\lim_{L\rightarrow\infty}r_{1}^{\frac{1}{2}(1+z)}\,,
\end{eqnarray}
\begin{eqnarray}
    \mathcal{F}_{C}
    &=&
    (1+z)l^{z-1}\frac{\Gamma(1+z/2)\zeta_{R}(2+z)}{4\pi^{z}\Gamma(-z/2)}\lim_{L\rightarrow\infty}\left[r_{1}^{\frac{1}{2}(z-1)}\left(\frac{1}{a^3}-\frac{3}{a^{4}}\int_0^{1}\tilde{h}du + \frac{6}{a^{5}}\int_0^{1}\tilde{h}^{2}du\right)\right]\,.
    \nonumber\\
\end{eqnarray}
If we consider roughness up to the second order in perturbations, the Casimir force density takes the form\footnote{Note that to evaluate the limit we need an explicit form for the periodic function $\tilde{h}$.}
\begin{eqnarray}
    \mathcal{F}_C
    &=&
    (1+z)l^{z-1}\frac{
    \Gamma\left(
    1+z/2
    \right)\zeta_R(2+z)
    }{4\pi \Gamma(-z/2)}
    \lim_{L\rightarrow\infty}\left[
    \frac{1}{a^{z+2}}
    - \frac{z+2}{a^{z+3}}\int_0^1\tilde{h}du
    \right.
    \nonumber
    \\
    &&
    \left.
    + \frac{3(z+3)}{2a^{z+4}}\int_0^1\tilde{h}^2du
    + \frac{(z-1)(z+3)}{2a^{z+4}}\left(
    \int_0^1\tilde{h}du
    \right)^2
    \right]
    \,.
    \label{FcLinf}
\end{eqnarray}
The unusual aspect in this result is the change in the orientation of the force each time we fixed different $z$ values \cite{Cheng:2022mwd, Borquez:2023cuf}. As we can see, independent of the roughness structure $\tilde{h}$, this change in the orientation of the force density $\mathcal{F}_C$ is determined exclusively by the anisotropic scaling factor present in the gamma function. To visualize this more clearly, we consider the case without roughness for values of $z=1, 3$, respectively,
\begin{equation}
    \mathcal{F}_C =
    - \frac{\zeta_R(3)}{8\pi a^3}\,,
    \qquad
    \mathcal{F}_C =
    \frac{9l^2\zeta_R(5)}{16\pi a^5}\,.
\end{equation}

\begin{figure}[H]
    \begin{minipage}[b]{0.8\linewidth}
    \centering
    \includegraphics[width=.6\linewidth]{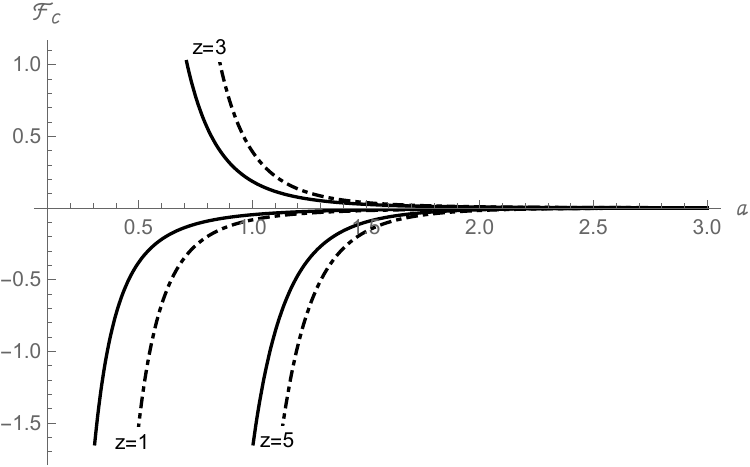} 
    \vspace{4ex}
  \end{minipage}
  \caption{ Casimir force density versus separation distance $a$. The solid curves represent the case without roughness ($h=0$) and the dashed curves represent the case with roughness ($h\neq 0$), both for different values of the anisotropic scaling factor $z=1,3,5$.}
  \label{fig1}
  \end{figure}

Now, we provide a specific example of a membrane with periodic roughness considering original coordinates of the form
\begin{eqnarray}
    h(x)=
    \epsilon\cos(\alpha  x+\theta)\,,
    \label{hcos}
\end{eqnarray}
where $\alpha$ has units of inverse length unit and $\theta$ is an arbitrary phase ($\epsilon$ represents a small parameter that indicates the perturbative nature of $h$). Taking into account this periodic function, the Casimir force density takes on the specific form
\begin{eqnarray}\label{FcEx1}
    \mathcal{F}_C
    =
    (1+z)l^{z-1}\frac{
    \Gamma\left(
    1+z/2
    \right)\zeta_R(2+z)
    }{4\pi \Gamma(-z/2)}
    \left(\frac{1}{a^{z+2}}
    + \frac{3(z+3)}{4}\frac{\epsilon^{2}}{a^{z+4}}\right)\,.
\end{eqnarray}
This result is a generalization of what was found in Ref. \cite{Droguett:2014gqa}. In Fig.\,\ref{fig1}, we can observe an increase in the magnitude of the force density due to the presence of additional contributions arising from the second order in perturbations, and to anisotropic scaling factor. As we commented above, for the cases of even anisotropic values, the force density is zero.


\section{Casimir effect at finite temperature}

To investigate the impact of temperature, it is convenient to employ the path integral approach considering imaginary time, which is associated with the finite temperature in the system. The path integral for the scalar field is given by
\begin{equation}
    Z=
    \int\mathcal{D}\phi\exp{(S(\phi))}\,.
\end{equation}
The effective action associated to operator $\mathcal{O}$ is 
\begin{equation}
    \Gamma=
    -\ln(Z)=
    \frac{1}{2}\ln\det[
    (
    -\partial^{2}_{\tau}
    +\mathcal{P}
    )/\mu
    ]=
    \frac{1}{2}\ln\det
    [
    \mathcal{O}/\mu
    ]\,,
\end{equation}
where $\mu$ is an arbitrary parameter with  mass dimension, introduced to render
the $\zeta$--function dimensionless. Eventually, the $\zeta$--function will be independent of this parameter $\mu$, hence we set $\mu=1$ for simplicity. Therefore, the Casimir energy is given by
\begin{eqnarray}
   E_{C}
   =
   \frac{\partial}{\partial\xi}\Gamma
   =
   \left.
   -\frac{1}{2}\frac{\partial}{\partial\xi}\left(\frac{d}{ds}\zeta_{\mathcal{O}}(s)
   \right)
   \right|_{s=0}
   \,,
\end{eqnarray}
where $\xi=1/T$ is the inverse of the  temperature.

The eigenvalue problem of the operator $\mathcal{O}$ is expressed by
\begin{eqnarray}
   (-
   \partial^{2}_{\tau}
   +(-1)^{z}l^{2(z-1)}\Delta^{z}
   )\phi
   =
   \omega\phi\,,
\end{eqnarray}
whose solution we propose for the anisotropic scalar field with $\tau\in\mathbb{C}$ is
\begin{eqnarray}
   \phi_{k,n,m}(\tau,x^i)=
   \frac{1}{\xi}e^{\frac{2\pi i k}{\xi}\tau}\varphi_{n,m}(x^i)
   \,,
\end{eqnarray}
where the eigenvalues from the time derivative are defined by periodic border, and correspond to $\omega_{k}=\frac{2\pi k}{\xi}$, with $k\in\mathbb{Z}$. Then, the total eigenvalues associated with the operator $\mathcal{O}$ are given by
\begin{eqnarray}
   \omega_{k,n,m}
   =
   \left(
   \frac{2\pi k}{\xi}\right)^{2}
   + l^{2(z-1)}
   \left(r_1n^2
        +r_2m^2
        \nonumber
        \right)^{z}\,
\end{eqnarray}
(Note that $r_{1}$ and $r_{2}$ are defined by (\ref{R12})). We use the integral representation of the $\zeta$--function to rewrite the spectral function as
\begin{eqnarray}\label{ZetaSum}
   \zeta_{\mathcal{O}}(s)
   =
   \frac{1}{\Gamma
   \left(
   s
   \right)}
   \int_{0}^{\infty}dt\,
   t^{s-1}
   \sum_{k=-\infty}^{\infty}
   \sum_{m,n=1}^{\infty}
   \exp\left\lbrace
   -t\left[
   \left(
   \frac{2\pi k}{\xi}
   \right)^{2}
   + l^{2(z-1)}
   \left(r_1n^2+
        r_2m^2
        \right)^z\right]\right\}\,.
   \nonumber\\
\end{eqnarray}
As was done in the case of zero temperature, it is possible to use Poisson resummation in (\ref{ZetaSum}),
\begin{eqnarray}
 \zeta_{\mathcal{O}}(s)
 &=&
 \frac{\xi}{\sqrt{4\pi}}
 \frac{\Gamma\left(
 s-1/2
 \right)}
 {\Gamma\left(s\right)}
\zeta_{\mathcal{P}}(s-1/2)
 \nonumber\\
   &&
   +\frac{\xi}{\sqrt{\pi}\Gamma{(s)}}
   \sum_{k,n,m=1}^\infty
   \int_0^\infty\,dt
   \,t^{s-3/2}
   \exp
   \left[
   -\frac{\xi^2 k^2}{4t}
   -l^{2(z-1)}\left(r_1n^2
        +r_2m^2
        \nonumber
        \right)^{z}t\right]\,.
        \nonumber\\
\end{eqnarray}
The energy is obtained by taking the derivative of the $\zeta$--function with respect to $s$ and evaluating it at $s=0$. Then, by integrating it, we obtain
\begin{eqnarray}
     \zeta_{\mathcal{O}}'(0)
    = 
    - \xi\zeta_{\mathcal{P}}(-1/2)
    +2\sum_{k,n,m=1}^\infty
   \frac{1}{k}\exp\left[ 
   -\xi k l^{z-1}\left(
   r_1n^2+r_2m^2
   \right)^{z/2}
   \right]\,.
\end{eqnarray}
The sum over $k$ can be explicitly performed using a geometric series, hence the Casimir energy is
\begin{eqnarray}
    E_{C}
    =
    \frac{1}{2}\zeta_{\mathcal{P}}(-1/2)
   + l^{z-1}\sum_{n,m=1}^{\infty} 
   \left[
   \frac{\left(
   r_1n^2
   +r_2m^2
   \right)^{z/2}}{\exp\left(
   \xi l^{z-1}\left(
   r_1n^2
   +r_2m^2
   \right)^{z/2}
   \right)-1}
   \right]
   \,.
\end{eqnarray}
In this result, it is  crucial to remember that we need to consider the finite terms of $\zeta_\mathcal{P} $ when we evaluated at $s=-1/2$ (see Eq. (\ref{zetaPs})).

\begin{figure}[H]
    \begin{minipage}[b]{0.5\linewidth}
    \centering
    \includegraphics[width=.83\linewidth]{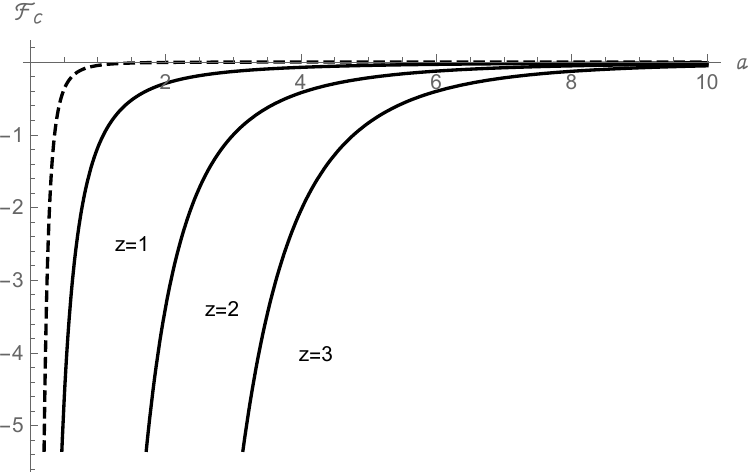} 
    \caption*{(a)}
    \vspace{4ex}
  \end{minipage}
    \begin{minipage}[b]{0.5\linewidth}
    \centering
    \includegraphics[width=.83\linewidth]{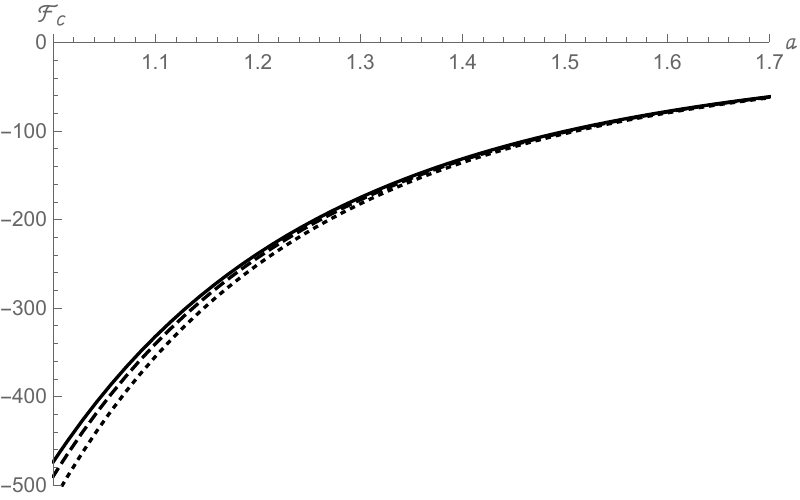}
    \caption*{(b)}
    \vspace{4ex}
  \end{minipage}
  \caption{Both graphs represent the Casimir force density versus separation distance. (a) The solid curves represent roughness for different values of the anisotropic scaling parameter $z=1,2,3$, with a constant temperature $T=100$. The dashed curve represents the case without roughness and $z=1$. (b) Different values of temperature are shown: solid curve for $T=100$, dashed curve for $T=150$, and dotted curve for $T=1000$, considering a constant value of the anisotropic scaling $z=3$.}
  \label{fig2}
  \end{figure}

The Casimir force density is obtained deriving the Casimir energy density with respect to the separation $a$.  Henceforth, we proceed with numerical calculations. In the Fig.\,\ref{fig2}, we present the force density versus  separation distance graphs for the case of a membrane with periodic border (see example in Eq. (\ref{hcos})), considering the effects of temperature and the anisotropic scaling factor. In the Fig.\,\ref{fig2} (a), an increase in the magnitude of the force is observed, which is compared to the case zero temperature. Likewise, it is noted that as the anisotropic scaling factor increases, the magnitude of the force density also increases. In contrast to the case zero temperature, there is a contribution to the Casimir force density for even values of the anisotropic scaling factor. In the Fig.\,\ref{fig2} (b), when the temperature increases for a constant value of the anisotropic scaling factor, there is an increase in the magnitude of the force. Therefore, the factors contributing to the intensification magnitude of the Casimir force density are the roughness, the anisotropic factor and temperature. 
  
\section{Conclusions}

We investigate  the case of a rough membrane embedded in a $2+1$--dimensional flat manifold. Vacuum fluctuations are induced by an anisotropic scalar field, whose Lagrangian is formulated within the framework of Ho\v rava-Lifshitz theory. Additionally, we  consider Dirichlet boundary conditions. To address the roughness, we perform a change of variables that leaves the edges of the membrane flat, and the remaining terms are considered as a perturbative roughness that is part of a potential. To determine the eigenvalues, we resort to perturbation theory up to first order and apply the $\zeta$--function regularization method.
We present an explicit case of a membrane with periodic border. In the scenario zero temperature, both the Casimir energy and force density depend on the anisotropic scaling factor. For certain values of this anisotropic factor, the energy and force density change orientation and, particularly, in cases where is even, both the energy and force density tend to zero. Moreover, the contribution of roughness arises from the second order in perturbations, increasing the magnitude of the Casimir effect. When we consider the effects of temperature, a significant increase in the magnitude of the force density is observed. Therefore, temperature, roughness and the anisotropic scaling factor contribute to the increase in the magnitude of the Casimir energy and force density.


\section*{Acknowledgements}
   Interuniversitario de Iniciación en Investigación Asociativa project: IUP22-36.



\begin{thebibliography}{99}	

\bibitem{Casimir:1948dh}
H.~B.~G.~Casimir,
\emph{On the Attraction Between Two Perfectly Conducting Plates},
Indag. Math. \textbf{10}, 261-263 (1948).

\bibitem{Boyer:1968uf}
T.~H.~Boyer,
\emph{Quantum electromagnetic zero point energy of a conducting spherical shell and the Casimir model for a charged particle},
Phys. Rev. \textbf{174}, 1764-1774 (1968)
doi:10.1103/PhysRev.174.1764.

\bibitem{Lamoreaux:1996wh}
S.~K.~Lamoreaux,
\emph{Demonstration of the Casimir force in the 0.6 to 6 micrometers range},
Phys. Rev. Lett. \textbf{78}, 5-8 (1997)
[erratum: Phys. Rev. Lett. \textbf{81}, 5475-5476 (1998)]
doi:10.1103/PhysRevLett.78.5.

\bibitem{Bressi:2002fr}
G.~Bressi, G.~Carugno, R.~Onofrio and G.~Ruoso,
\emph{Measurement of the Casimir force between parallel metallic surfaces},
Phys. Rev. Lett. \textbf{88}, 041804 (2002)
doi:10.1103/PhysRevLett.88.041804
[arXiv:quant-ph/0203002 [quant-ph]].

\bibitem{Bordag:2001qi}
M.~Bordag, U.~Mohideen and V.~M.~Mostepanenko,
\emph{New developments in the Casimir effect},
Phys. Rept. \textbf{353}, 1-205 (2001)
doi:10.1016/S0370-1573(01)00015-1
[arXiv:quant-ph/0106045 [quant-ph]].

\bibitem{Teo:2011kt}
L.~P.~Teo,
\emph{Casimir interaction between a cylinder and a plate at finite temperature: Exact results and comparison to proximity force approximation},
Phys. Rev. D \textbf{84}, 025022 (2011)
doi:10.1103/PhysRevD.84.025022
[arXiv:1106.1251 [quant-ph]].

\bibitem{Zhao:2006rr}
Y.~Zhao, C.~G.~Shao and J.~Luo,
\emph{Finite temperature Casimir effect for corrugated plates},
Chin. Phys. Lett. \textbf{23}, 2928-2931 (2006)
doi:10.1088/0256-307X/23/11/013.

\bibitem{Bellucci:2019ybj}
S.~Bellucci, I.~Brevik, A.~A.~Saharian and H.~G.~Sargsyan,
\emph{The Casimir effect for fermionic currents in conical rings with applications to graphene ribbons},
Eur. Phys. J. C \textbf{80}, 281 (2020)
doi:10.1140/epjc/s10052-020-7819-8
[arXiv:1912.09143 [hep-th]].

\bibitem{Kirsten:2010zp}
K.~Kirsten,
\emph{Basic zeta functions and some applications in physics},
MSRI Publ. \textbf{57}, 101-143 (2010)
[arXiv:1005.2389 [hep-th]].

\bibitem{Anselmi:2008bq}
D.~Anselmi,
\emph{Weighted power counting and Lorentz violating gauge theories. I. General properties},
Annals Phys. \textbf{324}, 874 (2009)
doi:10.1016/j.aop.2008.12.005
[arXiv:0808.3470 [hep-th]].

\bibitem{Anselmi:2008bs}
	D.~Anselmi,
	\emph{Weighted power counting and Lorentz violating gauge theories. II. Classification},
	Annals Phys. {\bf 324} 1058 (2009) 
	[arXiv:0808.3474 [hep-th]].

  \bibitem{Ferrari:2010dj}
A.~F.~Ferrari, H.~O.~Girotti, M.~Gomes, A.~Y.~Petrov and A.~J.~da Silva,
\emph{Ho\v rava-Lifshitz modifications of the Casimir effect},
Mod. Phys. Lett. A \textbf{28}, 1350052 (2013)
doi:10.1142/S0217732313500521
[arXiv:1006.1635 [hep-th]].

\bibitem{MoralesUlion:2015tve}
I.~J.~Morales Ulion, E.~R.~Bezerra de Mello and A.~Y.~Petrov,
\emph{Casimir effect in Ho\v rava\textendash{}Lifshitz-like theories},
Int. J. Mod. Phys. A \textbf{30}, no.36, 1550220 (2015)
doi:10.1142/S0217751X15502206
[arXiv:1511.00489 [hep-th]].

\bibitem{Muniz:2014dga}
C.~R.~Muniz, V.~B.~Bezerra and M.~S.~Cunha,
\emph{Casimir effect in the Ho\v{r}ava--Lifshitz gravity with a cosmological constant},
Annals Phys. \textbf{359}, 55-63 (2015)
doi:10.1016/j.aop.2015.04.014
[arXiv:1405.5424 [hep-th]].

\bibitem{daSilva:2019iwn}
D.~R.~da Silva, M.~B.~Cruz and E.~R.~Bezerra de Mello,
\emph{Fermionic Casimir effect in Ho\v rava\textendash{}Lifshitz theories},
Int. J. Mod. Phys. A \textbf{34}, no.20, 1950107 (2019)
doi:10.1142/S0217751X19501070
[arXiv:1905.01295 [hep-th]].

\bibitem{Erdas:2023wzy}
A.~Erdas,
\emph{Magnetic corrections to the fermionic Casimir effect in Horava-Lifshitz theories},
Int. J. Mod. Phys. A \textbf{38}, no.22n23, 2350117 (2023)
doi:10.1142/S0217751X23501178
[arXiv:2307.06228 [hep-th]].

\bibitem{Cruz:2017kfo}
M.~B.~Cruz, E.~R.~Bezerra de Mello and A.~Y.~Petrov,
\emph{Casimir effects in Lorentz-violating scalar field theory},
Phys. Rev. D \textbf{96}, no.4, 045019 (2017)
doi:10.1103/PhysRevD.96.045019
[arXiv:1705.03331 [hep-th]].

\bibitem{Erdas:2020ilo}
A.~Erdas,
\emph{Casimir effect of a Lorentz-violating scalar in magnetic field},
Int. J. Mod. Phys. A \textbf{35}, no.31, 2050209 (2020)
doi:10.1142/S0217751X20502097
[arXiv:2005.07830 [hep-th]].

\bibitem{Cruz:2018bqt}
M.~B.~Cruz, E.~R.~Bezerra De Mello and A.~Y.~Petrov,
\emph{Thermal corrections to the Casimir energy in a Lorentz-breaking scalar field theory},
Mod. Phys. Lett. A \textbf{33}, no.20, 1850115 (2018)
doi:10.1142/S0217732318501158
[arXiv:1803.07446 [hep-th]].

\bibitem{Erdas:2021xvv}
A.~Erdas,
\emph{Thermal effects on the Casimir energy of a Lorentz-violating scalar in magnetic field},
Int. J. Mod. Phys. A \textbf{36}, no.20, 20 (2021)
doi:10.1142/S0217751X21501554
[arXiv:2103.12823 [hep-th]].

\bibitem{Cheng:2022mwd}
H.~Cheng,
\emph{The Ho\v rava\textendash{}Lifshitz modifications of the Casimir effect at finite temperature revisited},
Eur. Phys. J. C \textbf{82}, no.11, 1032 (2022)
doi:10.1140/epjc/s10052-022-10854-4
[arXiv:2209.14544 [hep-th]].

\bibitem{Borquez:2023cuf}
C.~B\'orquez and B.~Droguett,
\emph{Casimir effect in 2+1 Ho\v{r}ava gravity},
Phys. Lett. B \textbf{844}, 138096 (2023)
doi:10.1016/j.physletb.2023.138096
[arXiv:2301.04566 [hep-th]].

\bibitem{Horava:2009uw}
  P.~Ho\v{r}ava,
  \emph{Quantum Gravity at a Lifshitz Point},
  Phys.\ Rev.\  D {\bf 79} (2009) 084008
  [arXiv:0901.3775 [hep-th]].
  
\bibitem{Horava:2008ih}
  P.~Ho\v{r}ava,
  \emph{Membranes at Quantum Criticality},
  JHEP {\bf 0903} (2009) 020
  [arXiv:0812.4287 [hep-th]].

  \bibitem{Blas:2009qj}
D.~Blas, O.~Pujolas and S.~Sibiryakov,
\emph{Consistent Extension of Ho\v rava Gravity},
Phys. Rev. Lett. \textbf{104}, 181302 (2010)
doi:10.1103/PhysRevLett.104.181302
[arXiv:0909.3525 [hep-th]].

\bibitem{Bellorin:2019zho}
J.~Bellor\'\i{}n and B.~Droguett,
\emph{Point-particle solution and the asymptotic flatness in 2+1D Ho\v{r}ava gravity},
Phys. Rev. D \textbf{100}, 064021 (2019)
doi:10.1103/PhysRevD.100.064021
[arXiv:1905.02836 [gr-qc]].

\bibitem{Arnowitt:1962hi}
R.~L.~Arnowitt, S.~Deser and C.~W.~Misner,
\emph{The Dynamics of general relativity},
Gen. Rel. Grav. \textbf{40}, 1997-2027 (2008)
doi:10.1007/s10714-008-0661-1
[arXiv:gr-qc/0405109 [gr-qc]].


\bibitem{Kiani:2012gh}
B.~Kiani and J.~Sarabadani,
\emph{Repulsive Casimir interaction between conducting and permeable corrugated plates},
Phys. Rev. A \textbf{86}, 022516 (2012)
doi:10.1103/PhysRevA.86.022516.

\bibitem{Setare:2014eva}
M.~R.~Setare and A.~Seyedzahedi,
\emph{Lateral Casimir Force between Two Sinusoidally Corrugated Eccentric Cylinders Using Proximity Force Approximation},
Acta Phys. Polon. B \textbf{45}, no.5, 1119 (2014)
doi:10.5506/APhysPolB.45.1119
[arXiv:1402.3652 [hep-th]].

\bibitem{Setare:2013vba}
M.~R.~Setare and A.~Seyedzahedi,
\emph{Casimir Energy between a Sinusoidally Corrugated Sphere and a Plate Using Proximity Force Approximation},
Indian J. Phys. \textbf{90}, no.5, 583-588 (2016)
doi:10.1007/s12648-015-0781-x
[arXiv:1311.4022 [cond-mat.other]].

\bibitem{Droguett:2014gqa}
B.~Droguett and J.~C.~Rojas,
\emph{Casimir energy of non-flat bordered membrane},
Mod. Phys. Lett. A \textbf{29}, 1450127 (2014)
doi:10.1142/S0217732314501272.

\end{thebibliography}
\end{document}